\documentclass[preprint2]{aastex6}
\usepackage{txfonts}
\usepackage{upgreek}
\usepackage{graphicx}
\newcommand*{\unitstyle}[1]{\mathrm{#1}}
\newcommand*{\unitskip}{\,}                             %
\newcommand{\power}[2]{\ensuremath{{#1}^{#2}}}		    

\newcommand*{\meter}{\unitstyle{m}}

\newcommand*{\second}{\unitstyle{s}}

\newcommand*{\centi}{\unitstyle{c}}		

\newcommand*{\cm}{\centi\meter}
\newcommand*{\gram}{\unitstyle{g}}

\newcommand*{\grampercc}{\gram\unitskip\power{\cm}{-3}} 
\newcommand*{\grampersquarecm}{\gram\unitskip\power{\cm}{-2}} 

\newcommand{\EE}[2]{\ensuremath{{#1}\times 10^{#2}}}
\newcommand{\density}[2]{$\rho={#1}\times 10^{#2}\,\grampercc$}

\newcommand*{\mdot}{\dot{m}}
\newcommand*{\mdotEdd}{\mdot_{\mathrm{Edd}}}

\begin{document}
\title{The Impact of Neutron Transfer Reactions on Heating and Cooling of Accreted Neutron Star Crusts }

\email[Correspondence to: ]{schatz@nscl.msu.edu}

\author{
H. Schatz\altaffilmark{1,2,3},
Z. Meisel\altaffilmark{3,5},
E. F. Brown\altaffilmark{1,2,3,4},
S. S. Gupta\altaffilmark{6},
G. W. Hitt\altaffilmark{7},
W.~R. Hix\altaffilmark{8,9},
R. Jain\altaffilmark{1,2,3},
R. Lau \altaffilmark{10},
P.  M{\"o}ller\altaffilmark{3},
W.-J. Ong\altaffilmark{3,11},
P. S. Shternin\altaffilmark{12},
Y. Xu\altaffilmark{13},
M. Wiescher\altaffilmark{3,14} 
}

\altaffiltext{1}{National Superconducting Cyclotron Laboratory, Michigan State University, 640 South Shaw Lane, East Lansing, Michigan 48824, USA.}
\altaffiltext{2}{Department of Physics and Astronomy,Michigan State University, 567 Wilson Road, East Lansing, Michigan 48824, USA.}
\altaffiltext{3}{Joint Institute for Nuclear Astrophysics, Center for the Evolution of the Elements.}
\altaffiltext{4}{Department of Computational Mathematics, Science and Engineering, Michigan State University, East Lansing, Michigan 48824, USA.}
\altaffiltext{5}{Institute of Nuclear \& Particle Physics, Department of Physics \& Astronomy, Ohio University, Athens, Ohio 45701, USA.}
\altaffiltext{6}{Indian Institute of Technology Ropar, Nangal Road, Rupnagar (Ropar), Punjab 140 001, India.}
\altaffiltext{7}{Department of Physics and Engineering Science, Coastal Carolina University, P.O. Box 261954 Conway, SC 29528, USA.}
\altaffiltext{8}{Physics Division, Oak Ridge National Laboratory, PO Box 2008, Oak Ridge, Tennessee 37831-6354, USA.}
\altaffiltext{9}{Department of Physics and Astronomy, University of Tennessee, 401 Nielsen Physics Building, 1408 Circle Drive, Knoxville, Tennessee 37996-1200, USA.}
\altaffiltext{10}{HKU SPACE, University of Hong Kong, Pok Fu Lam, Hong Kong.}
\altaffiltext{11}{Nuclear and Chemical Sciences Division, Lawrence Livermore National Laboratory, Livermore, California 94551, USA}
\altaffiltext{12}{Ioffe Institute, Politekhnicheskaya 26, Saint Petersburg, 194021, Russia.}
\altaffiltext{13}{Extreme Light Infrastructure - Nuclear Physics (ELI-NP), Horia Hulubei National Institute for R\&D in Physics and Nuclear Engineering (IFIN-HH), 077125 Buchurest-Magurele, Romania.}
\altaffiltext{14}{Department of Physics, 225 Nieuwland Science Hall, University of Notre Dame, Notre Dame, Indiana 46556, USA.}

\begin{abstract}
Nuclear reactions heat and cool the crust of accreting neutron stars and need to be understood to interpret observations of X-ray bursts and of long-term cooling in transiently accreting systems. It was recently suggested that previously neglected neutron transfer reactions may play a significant role in the nuclear processes. We present results from full nuclear network calculations that now include these reactions and determine their impact on crust composition, crust impurity, heating, and cooling. We find that a large number of neutron transfer reactions indeed occur and impact crust models. In particular, we identify a new type of reaction cycle that brings a pair of nuclei across the nuclear chart into equilibrium via alternating neutron capture and neutron release, interspersed with a neutron transfer. While neutron transfer reactions lead to changes in crust model predictions, and need to be considered in future studies, previous conclusions concerning heating, cooling, and compositional evolution are remarkably robust.
\end{abstract}
\maketitle

\section{Introduction}
The crusts of transiently accreting neutron stars in X-ray binary systems cool over timescales of months during periods of quiescence, when the accretion turns off \citep{Rutledge2002,Cackett2006,Shternin2007,Brown2009}. For a sub-class of quasi-persistent transients the periods of quiescence can extend over years. In such systems, the decreasing surface temperature as a function of time can be inferred from repeated X-ray observations. These so called cooling curves have been shown to be interesting probes of the dense matter physics inside the neutron star crust, such as the lattice structure of the crust \citep{Cackett2006,Shternin2007,Brown2009}, compositional impurity of the crust \citep{Brown2009,Page2013,Turlione2015,Ootes2016,Merritt2016}, neutron superfluidity \citep{Shternin2007,Brown2009}, nuclear pasta \citep{Horowitz2015,Deibel2017}, and novel heating mechanisms at shallow depths \citep{Brown2009,Degenaar2011,Page2013,Degenaar2013,Degenaar2015,Deibel2015,Turlione2015,Waterhouse2016,Merritt2016,Waterhouse2016,Parikh2019,Chamel2020,Potekhin2021}. 

However, accurate interpretation of these observations requires crust models that capture the relevant nuclear physics processes during accretion \citep{Sato1979,Haensel1990,Haensel2008,Gupta2007,Gupta2008,Shternin2007,Schatz2014,Meisel2018,Lau2018,Shchechilin2019,Shchechilin2021}. These nuclear processes determine the location of nuclear heating and cooling and therefore the temperature profile at the beginning of the cooling phase. They also determine the composition of the crust as a function of depth, which directly affects the thermal transport properties, for example via the impurity parameter $Q_{\rm imp}=\sum_i Y_i (Z_i - \langle Z \rangle )^2 / \sum_i Y_i$ with charge numbers $Z_i$, average charge number $\langle Z \rangle$ and abundances $Y_i$ (excluding free neutrons). To that end, we recently carried out the first crust model calculations that use a full nuclear reaction network \citep{Lau2018}. The network included electron capture, $\beta$-decay, neutron capture, pycnonuclear fusion, and neutron emission. Neutron emission processes considered included thermal excitation (the inverse to neutron capture) as well as population of neutron unbound states via electron capture or $\beta$-decays. It was recently suggested that the proximity of the nuclei at the high densities in neutron star crusts can also lead to neutron transfer reactions, where a neutron can be transferred from a nucleus to its neighbor \citep{Chugunov2019}. Here we present results from updated reaction network calculations that include neutron transfer reactions. We demonstrate the impact of neutron transfer reactions on crust heating, cooling, and the composition of the accreted crust.  

\section{Model}
The crust model and nuclear reaction network are identical to \citet{Lau2018}, with the exception of the addition of neutron transfer reactions. We provide here a brief summary of the main features. To map the steady state compositional changes as a function of depth, we follow the composition of an accreted fluid element with time $t$ in a plane-parallel 1D approximation. Pressure rises as $P=\mdot gt$, with local accretion rate $\mdot=0.3\mdotEdd$ in the rest frame at the surface, local Eddington accretion rate $\mdotEdd=\EE{8.8}{4}\,\grampersquarecm\,\second^{-1}$, and surface gravity $g=\EE{1.85}{14}\,\cm\,\second^{-2}$. The temperature $T$ is kept constant at 0.5~GK, while the mass density $\rho$ increases according to the equation of state $P=P(T,\rho,Y_i)$ as described in \citet{Gupta2007}. The initial composition is determined by previous thermonuclear hydrogen, helium, and carbon burning near the surface and is expected to differ from system to system depending on accretion rate, accreted composition, and neutron star properties. We explore two possibilities: superburst ashes \citep{Keek2012} or the ashes of an extreme rapid proton capture process (rp-process) in a mixed hydrogen and helium X-ray burst \citep{Schatz2001}. Superburst  ashes consist mainly of nuclei in the $A=50-60$ mass range, while the rp-process  produces a broader range of nuclei in the $A=20-108$ range (see \citet{Lau2018} for details). 

Composition changes are followed with a nuclear reaction network that tracks the changes of abundances of individual nuclear species as a function of depth and provides nuclear heat release as well as cooling rates via neutrino emission. The nuclear physics input is identical to \citet{Lau2018} to facilitate comparison with the exception of the addition of neutron transfer reactions. 

Individual neutron transfer reactions are calculated using the approximation derived by \citet{Chugunov2019}. The approach assumes that the transition rate for neutron transfer obeys Fermi's golden rule. The nuclear density of states for the neutron acceptor nucleus is roughly $\varrho=1$~MeV$^{-1}$.  The neutron donor and acceptor nuclei overlap in the asymptotic region of the wave function that is well described by an exponential. Transfer will then occur for the least-bound neutron of the donor nucleus. Employing first order estimates for the wave function overlap as well as the volume average of the transition rate weighted by the number density of acceptor nuclei, the neutron transfer reaction  rate between neutron donor $\mathcal{D}$ with charge number $Z_{\cal D}$ and neutron acceptor $\mathcal{A}$ with charge number $Z_{\cal A}$ is approximated by
 \begin{equation}
     \label{eqn:transrate}
     \lambda_{\rm \mathcal{AD}}=4\pi\,n_{\rm \mathcal{A}}l_{\rm pk}^{3}\sqrt{\frac{\pi\,l_{\rm pk}}{a_{\rm \mathcal{AD}}\Gamma_{\rm \mathcal{AD}}}}W(l_{\rm pk})g_{\rm \mathcal{AD}}(l_{\rm pk}).
 \end{equation}
 Here, $n_{\mathcal{A}}=N_{\mathrm{A}}Y_{\mathcal{A}}\rho$ is the number density of $\mathcal{A}$, and $N_{\mathrm{A}}$ is the Avogadro constant. The average ion sphere radius is
 \begin{equation}
     \label{eqn:radius}
     a_{\mathcal{AD}}=\frac{1}{2}\left(Z_{\mathcal{A}}^{1/3}+Z_{\mathcal{D}}^{1/3}\right)\left(\frac{3}{4\pi n_{\mathrm{e}}}\right)^{1/3}
 \end{equation}
for electron density $n_{e}=N_{A}Y_{e}\rho$ at electron fraction $Y_{e}$. The Coulomb coupling parameter is
\begin{equation}
    \label{eqn:coupling}
    \Gamma_{\mathcal{AD}}=\frac{Z_{\mathcal{A}}Z_{\mathcal{D}}\alpha\hbar c}{a_{\mathcal{AD}}k_{\rm B}T}.
\end{equation}
with fine structure constant $\alpha$ and speed of light $c$. The neutron transfer rate depends on the distance between $\mathcal{A}$ and $\mathcal{D}$. The distance between $\mathcal{A}$ and $\mathcal{D}$ that provides the main contribution to the total neutron transfer reaction rate if one were to integrate over all $\mathcal{A}-\mathcal{D}$ separation distances is
\begin{equation}
    \label{eqn:lpk}
    l_{\rm pk}=a_{\mathcal{AD}}\left(\frac{25}{64}+2\frac{\kappaup a_{\mathcal{AD}}}{\Gamma_{\mathcal{AD}}}\right)^{-1/2}
\end{equation}
with $\kappaup=\sqrt{2m_{n}S_{n}}/\hbar$ for neutron mass $m_{n}$ and binding energy within $\mathcal{D}$ at the neutron separation energy $S_{n}$. The neutron transfer transition probability per unit time is
\begin{equation}
    \label{eqn:wpk}
    W(l_{\rm pk})=3\times10^{21}\left(\frac{50}{l_{\rm pk}}\right)^{2}\exp\left(-22.6\frac{l_{\rm pk}}{50}\sqrt{S_{\mathrm{n}}}\right)\,{\rm s}^{-1}
\end{equation}
using units of MeV for $S_{n}$ and fm for $l_{\rm pk}$.
Finally, the pair-correlation function between $\mathcal{A}$ and $\mathcal{D}$ is 
\begin{equation}
    \label{eqn:pair}
    g_{\mathcal{AD}}(l_{\mathrm{pk}})=\exp\left(-\Gamma_{\mathcal{AD}}\left\{\frac{a_{\mathcal{AD}}}{l_{\rm pk}}\right\}-u_{\mathcal{AD}}(l_{\rm pk})\right)
\end{equation}
using an approximate potential of mean strength
\begin{equation}
    \label{eqn:meanpot}
    u_{\mathcal{AD}}(l_{\rm pk})=1.25-\frac{25}{64}\frac{l_{\rm pk}}{a_{\mathcal{AD}}}.
\end{equation}

We include all possible neutron transfer reactions between nuclei in the network that have a positive Q-value and a reaction timescale faster than 10$^{12}$~s at the highest densities at the end of our calculation. We also include reverse reactions based on detailed balance \citep{Arnett1996}. For the calculation with initial superburst ashes, we use a 894 nuclei network up to Zn that now includes 120,928 neutron transfer reactions. For initial rp-process ashes the network includes 1400 nuclei up to Cd and 299,825 transfer reactions. 

\section{Results}
\subsection{Superburst Ashes}
\begin{figure}
\epsscale{1.15}
\plotone{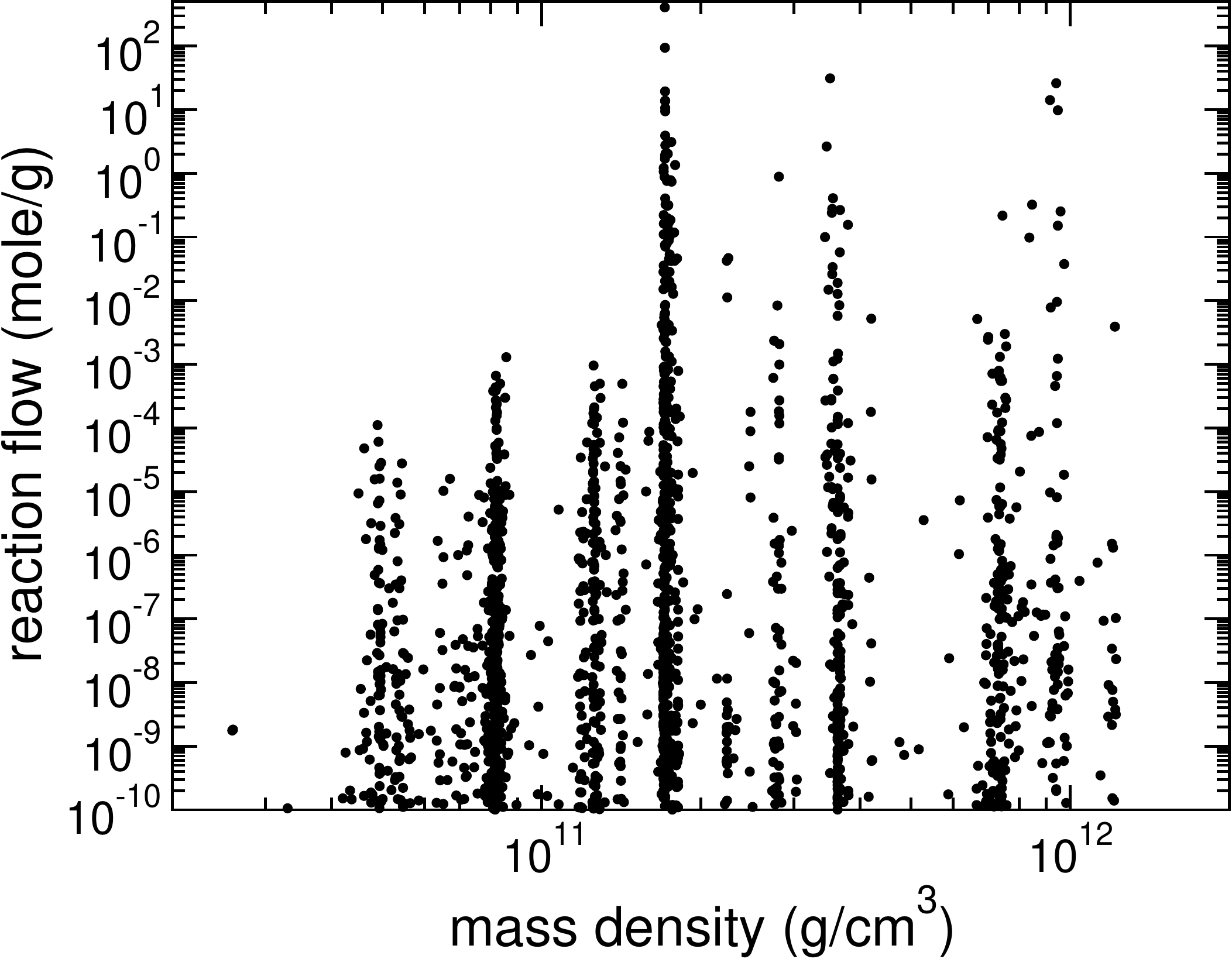}
\caption{\label{FigFluxes} Total time-integrated reaction flow through the dominant transfer reactions vs. the mass density where the respective reaction sets in (i.e. has achieved 10\% of its final integrated flow) for superburst ashes.}
\end{figure}

As we follow the composition of the initial superburst ashes with increasing density into the crust, the first significant neutron transfer reactions with time-integrated reaction flows $F$ above 10$^{-8}$~mole/g set in at a density of \density{4.6}{10}. Fig.~\ref{FigFluxes} shows the integrated reaction flows of the most dominant transfer reactions vs. the density where the reaction sets in. Already between \density{4.6}{10} and  \density{5.1}{10} there are 42 transfer reactions with $F > 10^{-8}$~mole/g. Up to \density{1.5}{12}, there are 1053 such reactions. At that depth, a significant free neutron abundance (about 25\%) is built up making neutron transfer at those and higher densities irrelevant. Fig.~\ref{FigFluxes} indicates that neutron transfer reactions cluster at certain densities. These are depths where electron capture thresholds of the most abundant species are reached and more neutron-rich nuclei with lower $S_{\mathrm{n}}$ are created. The newly created more neutron-rich nuclei may directly undergo a neutron transfer, or ($\gamma$,n) reactions may release neutrons that are in turn captured by other abundant nuclei, resulting in weakly bound nuclei with high neutron transfer rates. 

However, the time-integrated reaction flow is not necessarily a good indicator of the importance of a neutron transfer reaction. As Fig.~\ref{FigFluxes} shows, many of the reaction flows exceed the abundances of even the most abundant species (10$^{-4}$~mole/g). This indicates the presence of reaction cycles that can lead to large reaction flows without significant changes in composition. Indeed, many neutron transfer reactions become embedded in (n,$\gamma$)$-$($\gamma$,n) equilibria leading to such reaction cycles. An example is the $^{55}$Ca($^{22}$O,$^{23}$O)$^{54}$Ca neutron transfer reaction, which is the reaction with the largest time-integrated flow of 400~mole/g. This reaction sets in at \density{1.7}{11} where there is a high abundance of $^{54}$Ca and some $^{22}$O. The initially present $A=55$ nuclei have long been destroyed by other neutron transfer and ($\gamma$,n) reactions. However, due to early release of neutrons via ($\gamma$,n) reactions prior to neutron drip \citep{Lau2018} there is a significant $^{54}$Ca(n,$\gamma$)$^{55}$Ca reaction flow. The subsequent neutron transfer from $^{55}$Ca to $^{22}$O results in $^{54}$Ca and $^{23}$O. A $^{23}$O($\gamma$,n)$^{22}$O reaction then releases the transferred neutron and restores the initial composition, closing the cycle. This is effectively a (n,$\gamma$)$-$($\gamma$,n) equilibrium cycle where the neutron is captured by $^{54}$Ca but is released by $^{23}$O due to the intermediate neutron transfer step. 

\begin{figure}
\epsscale{1.15}
\plotone{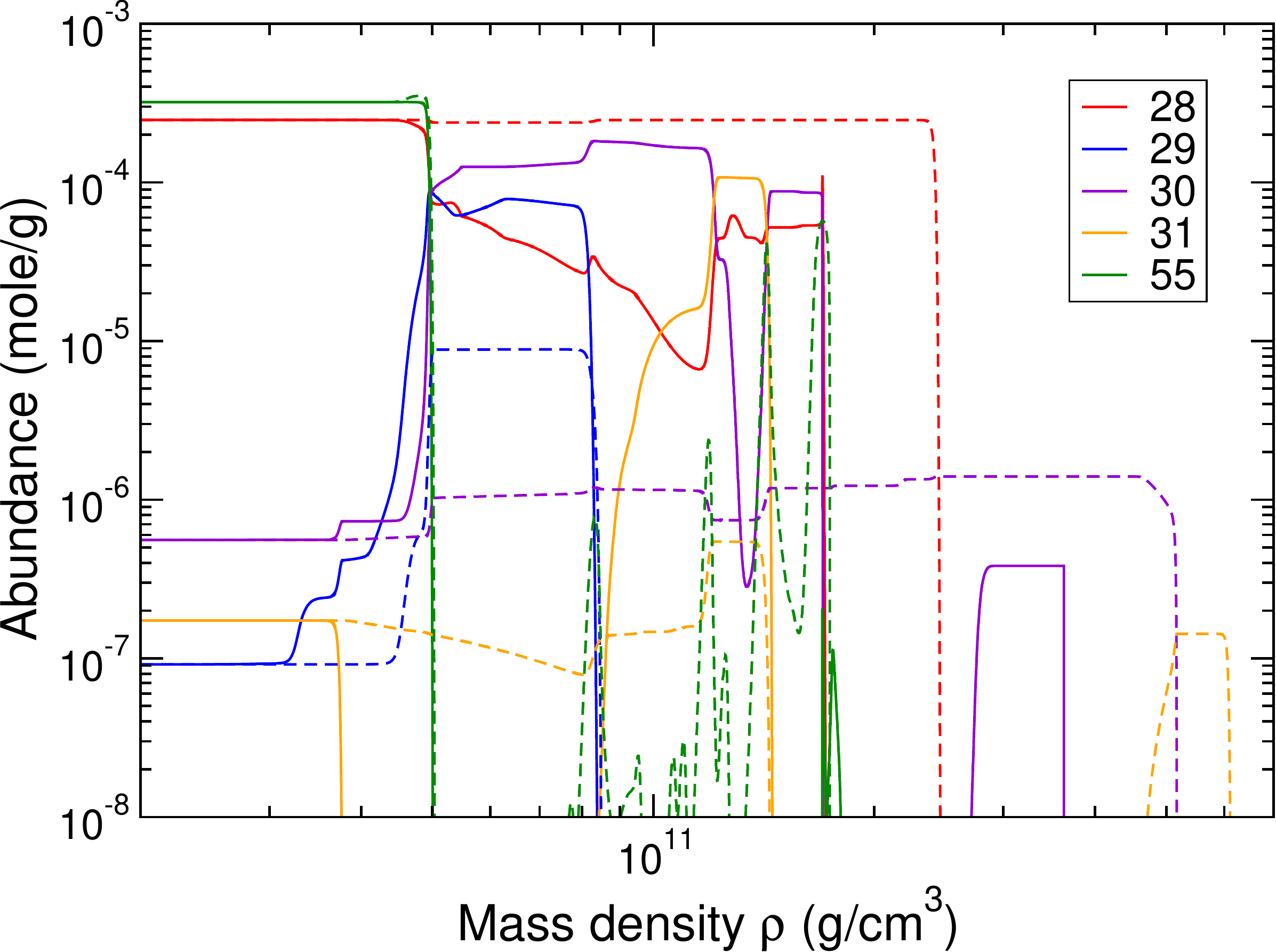}
\caption{\label{FigAbtA} Abundances for superburst ashes,  summed by mass number $A$ for $A=28$, 29, 30, 31, 55 indicated by the color, as functions of mass density, with neutron transfer reactions (solid) or without (dashed).}
\end{figure}

\begin{figure}
\epsscale{1.15}
\plotone{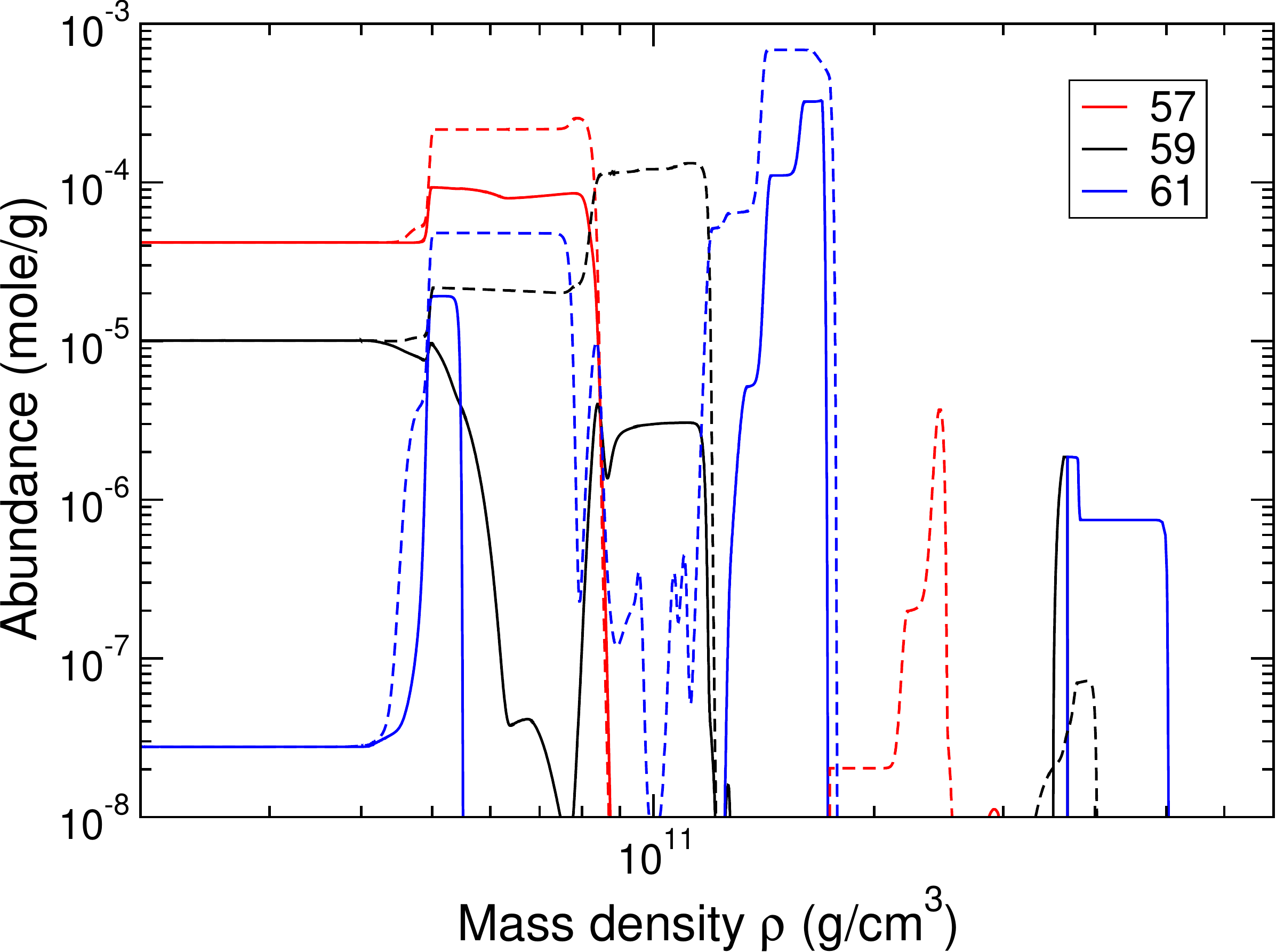}
\caption{\label{FigAbtB} Abundances for superburst ashes, summed by mass number $A$ for $A=57$, 59, 61 indicated by the color, as functions of mass density, with neutron transfer reactions (solid) or without (dashed).}
\end{figure}

To identify the important neutron transfer reactions that affect the neutron star crust composition significantly, we follow the mass numbers $A$ with the largest summed abundance $Y(A)$ up to \density{1.5}{12} and compare calculations with and without neutron transfer. Initially, at lower densities, electron captures dominate the composition changes, leaving the initial $Y(A)$ distribution unchanged. As neutron reactions set in, either transfer, emission, or capture, abundances are transferred between mass chains and new mass numbers can become abundant, while others are depleted. 
The initial composition is dominated (abundance larger than 10$^{-4}$ mole/g) by $^{28}$Si, $^{52}$Cr, $^{54}$Cr, $^{55}$Fe, $^{56}$Fe, $^{58}$Fe, and $^{60}$Ni. $Y(A)$ for $A=24, 26, 29, 30, 31,$ and 61 become significant deeper in the crust due to (n,$\gamma$), ($\gamma$,n), and neutron transfer reactions. Without transfer reactions, $A=24, 26, 30$, and 31 are never abundant, instead, $A=57, 59$ become more important, indicating already significant changes due to transfer reactions. Figs.~\ref{FigAbtA} and \ref{FigAbtB} show the abundances as functions of density for those significant $Y(A)$ that show the largest differences due to transfer reactions. As noted above, important changes start to occur at \density{4.6}{10}, where $A=55$ nuclei are destroyed. This destruction is similar with or without neutron transfer reactions, as it is dominated by destruction of the weakly neutron-bound $^{55}$Ca ($S_{n}=1.3$~MeV) via ($\gamma$,n) reactions. However, the $^{55}$Ca($^{28}$Mg,$^{29}$Mg)$^{54}$Ca neutron transfer reaction contributes and leads to a stronger buildup of $A=29$ and a coinciding drop in $A=28$. Without neutron transfer reactions, there is also a buildup of $^{29}$Mg via $^{28}$Mg(n,$\gamma$)$^{29}$Mg, but it is much smaller because the neutrons from $^{55}$Ca($\gamma$,n) are also captured by other species. As $^{29}$Mg is produced, the $^{29}$Mg($^{29}$Mg,$^{28}$Mg)$^{30}$Mg neutron transfer reaction becomes important at the same depth, leading to the strong buildup of $A=30$ that is only seen with transfer reactions included. The rapid depletion of the initial (though relatively small) $A=31$ abundance is due to the 
$^{31}$Mg($^{28}$Mg,$^{29}$Mg)$^{30}$Mg and $^{31}$Mg($^{29}$Mg,$^{28}$Mg)$^{31}$Mg neutron transfer reactions, while the destruction of $A=59$ is due to $^{59}$Ti($^{28}$Mg,$^{29}$Mg)$^{58}$Ti. In addition, transfer reactions significantly reduce the production of $A=57$ via $^{56}$Ti(n,$\gamma$)$^{57}$Ti and $A=61$ via $^{60}$Cr(n,$\gamma$)$^{61}$Cr. This is due to the smaller number of free neutrons available for neutron capture, as some of the $^{55}$Ca destruction is now due to neutron transfer instead of ($\gamma$,n), and only the latter produces free neutrons. 

Neutron transfer reactions lead to some additional compositional changes at higher densities. There is a strong build up of $A=31$ over a broad range of densities (Fig.~\ref{FigAbtA}), with the dominant part occurring around \density{1.2}{11}. This is in roughly equal parts due to the 
$^{33}$Mg($^{30}$Mg,$^{31}$Mg)$^{32}$Mg and  $^{30}$Na($^{30}$Mg,$^{31}$Mg)$^{29}$Na neutron transfer reactions. There is also some $A=26$ production, not shown in Fig.~\ref{FigAbtA}, mainly at around \density{1.3}{11} due to $^{28}$Ne($\gamma$,n)$^{27}$Ne followed by  $^{27}$Ne($^{31}$Na,$^{30}$Na)$^{26}$Ne.

\begin{figure}
\epsscale{1.15}
\plotone{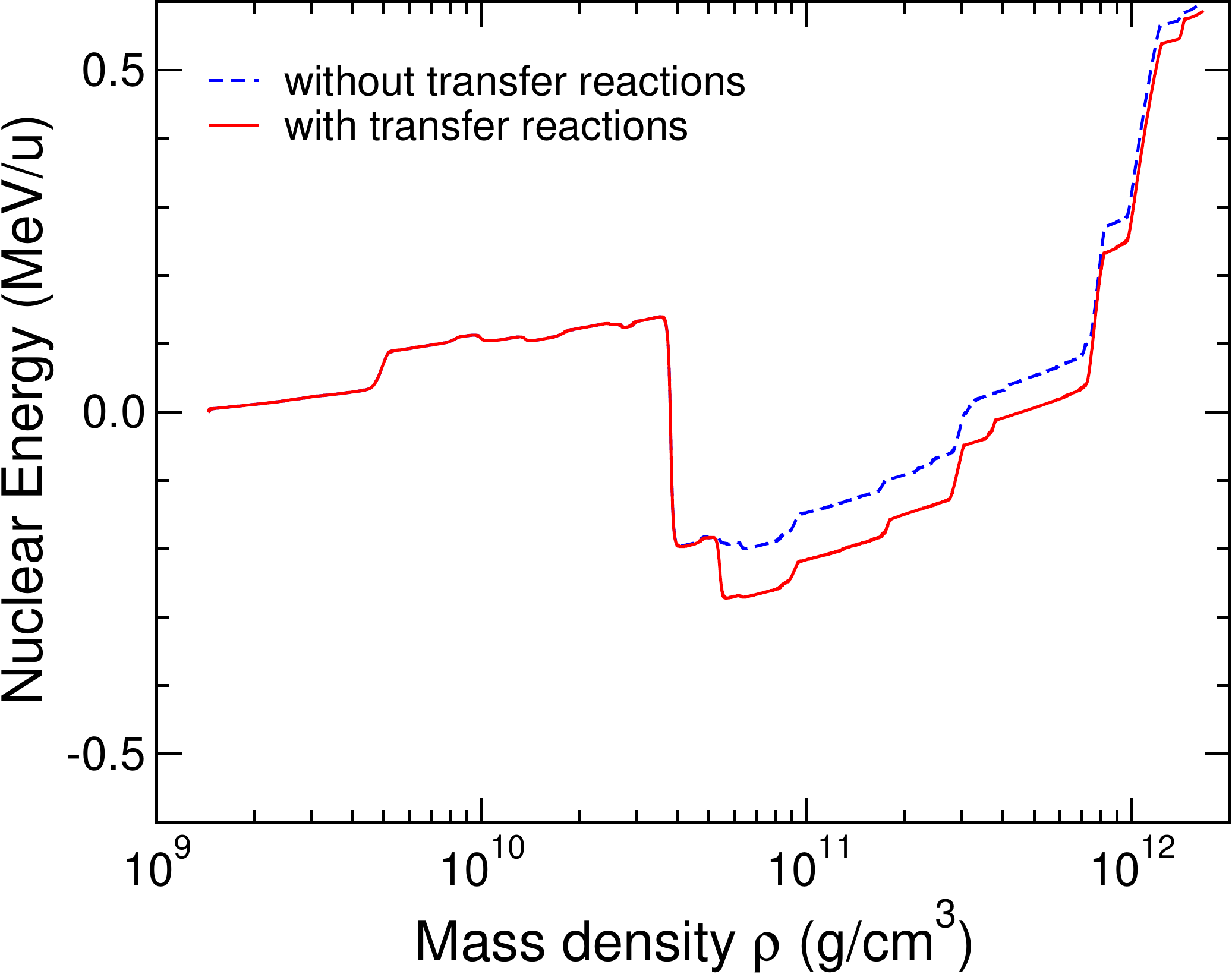}
\caption{\label{FigSbEnergy} Integrated nuclear energy deposited in the crust as a function of mass density for superburst ashes with neutron transfer reactions (red, solid) and without (blue, dashed).}
\end{figure}

\begin{figure}
\epsscale{1.15}
\plotone{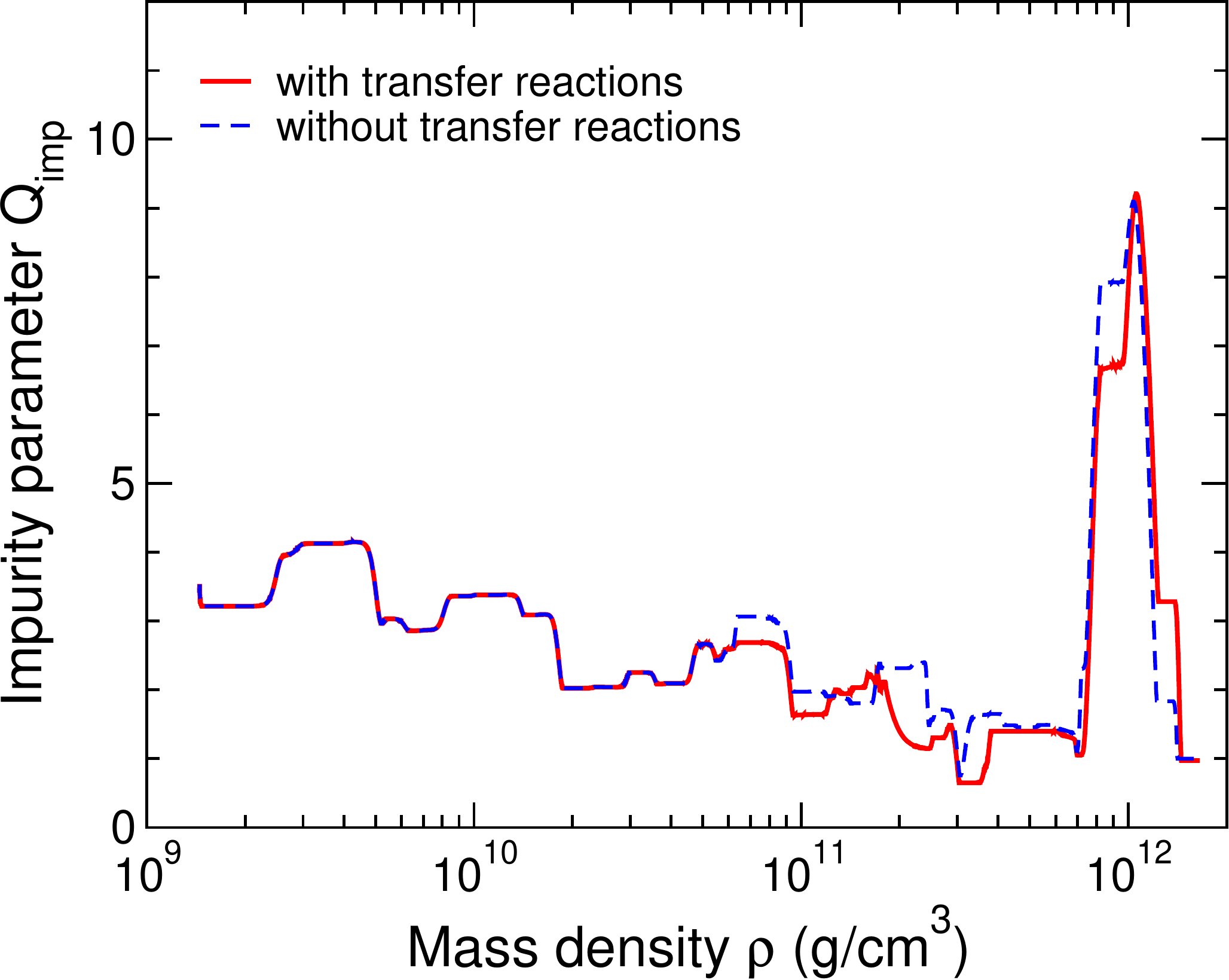}
\caption{\label{FigSbImpurity} Impurity parameter $Q_\mathrm{imp}$ (see text) as a function of mass density for superburst ashes with neutron transfer reactions (red, solid) and without (blue, dashed).}
\end{figure}

The integrated nuclear energy generation and the impurity parameter $Q_\mathrm{imp}$ as functions of density are shown in Fig.~\ref{FigSbEnergy}  and Fig.~\ref{FigSbImpurity}, respectively. The results without neutron transfer reactions are reproduced from \citet{Lau2018}. The negative slope in the integrated nuclear energy is caused by urca cooling \citep{Schatz2014,Deibel2016,Lau2018}. The slight increase in urca cooling with neutron transfer reactions stems from the increased $A=29$ abundance, which results in increased cooling from the $^{29}$Mg$\leftrightarrow ^{29}$Na urca pair. 

\subsection{rp-process ashes}
\begin{figure}
\epsscale{1.15}
\plotone{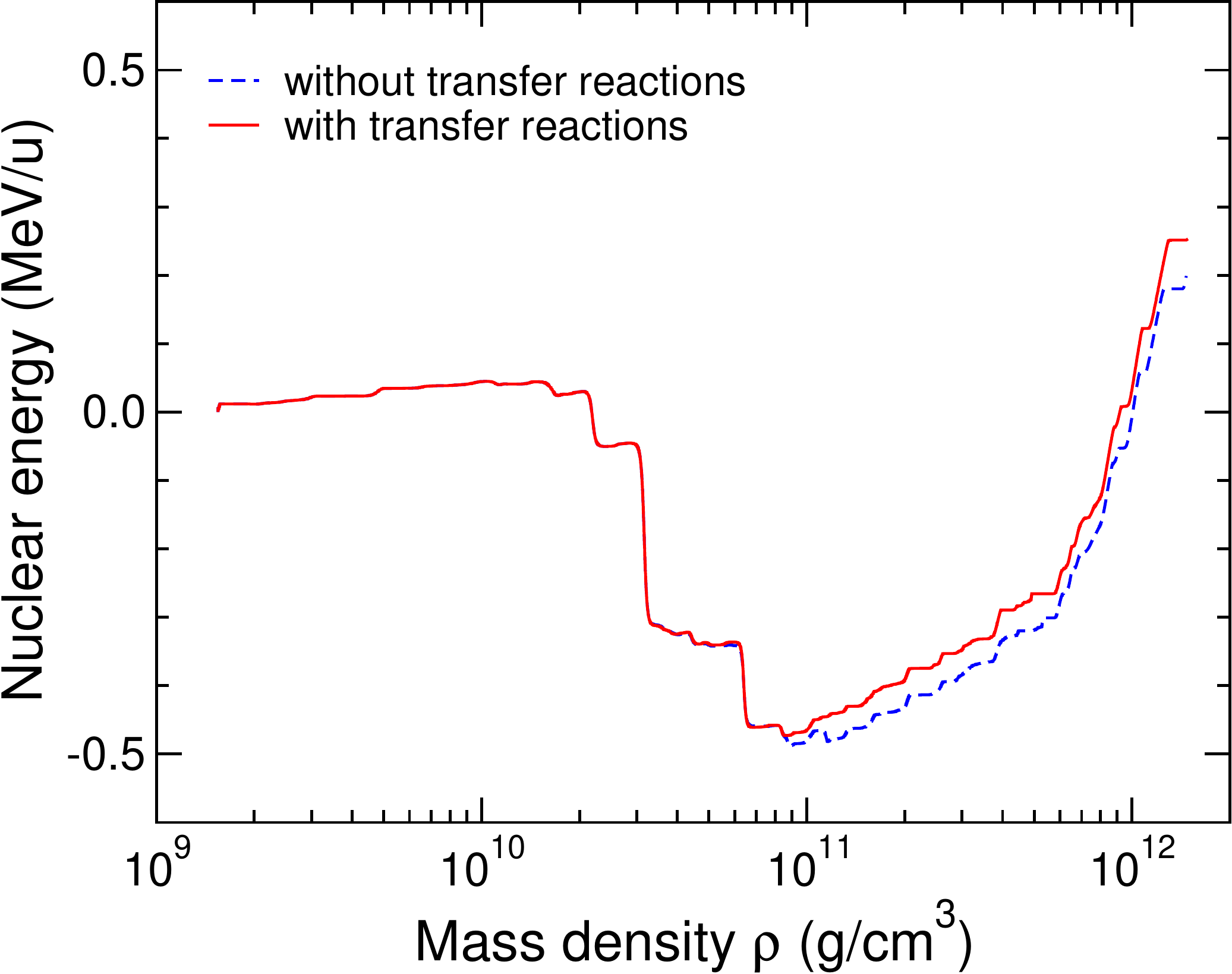}
\caption{\label{FigXrbEnergy} Integrated nuclear energy deposited in the crust as a function of mass density for rp-process ashes with neutron transfer reactions (red, solid) and without (blue, dashed).}
\end{figure}

\begin{figure}
\epsscale{1.15}
\plotone{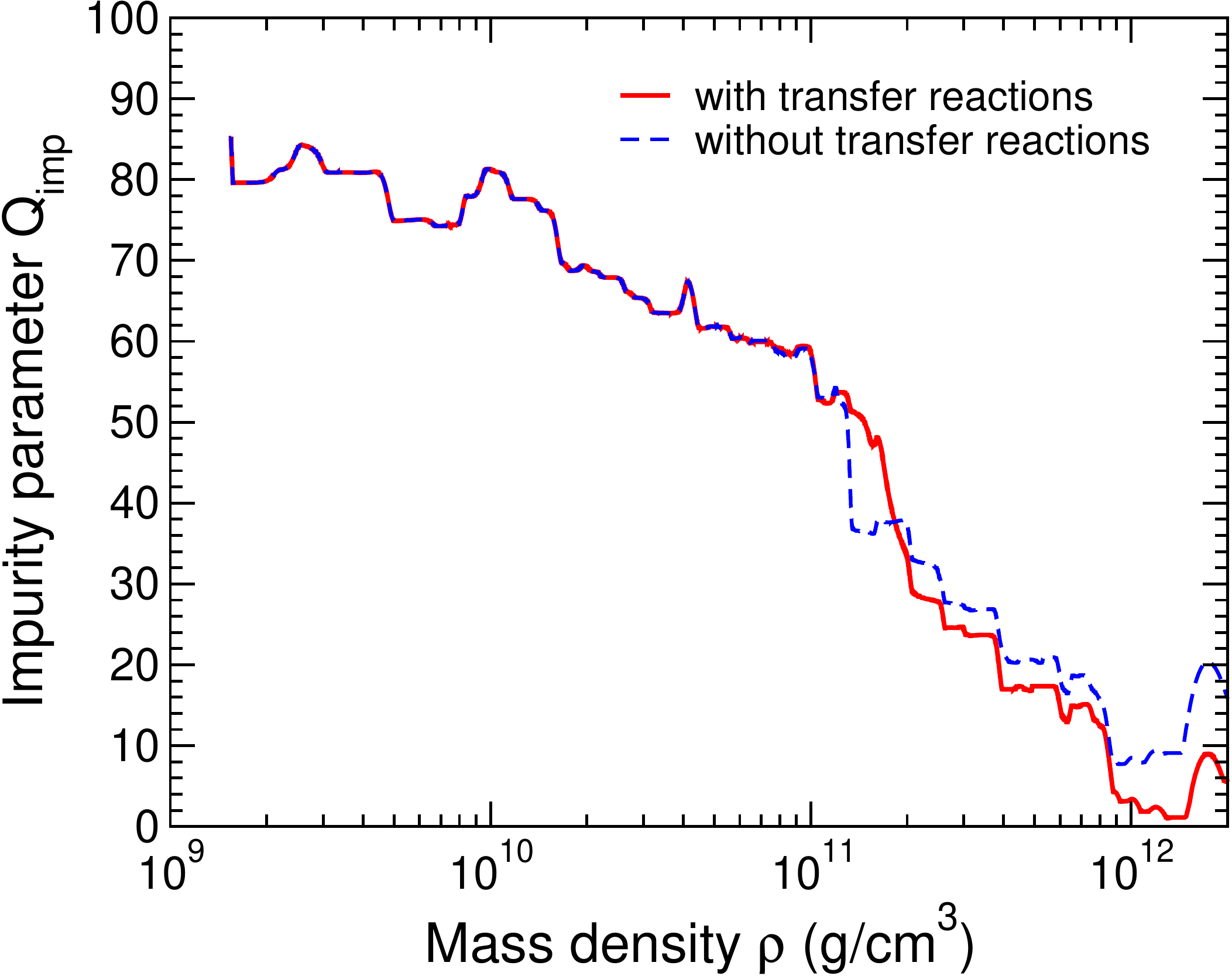}
\caption{\label{FigXrbImpurity} Impurity parameter $Q_\mathrm{imp}$ (see text) as a function of mass density for rp-process ashes with neutron transfer reactions (red, solid) and without (blue, dashed).}
\end{figure}

The model with initial rp-process ashes is characterized by a much broader initial composition range $20 \leq A \leq 108$. Again a large number of neutron transfer reactions occur. The integrated nuclear energy generation and the impurity parameter $Q_\mathrm{imp}$ as functions of density are shown in Fig.~\ref{FigXrbEnergy}  and Fig.~\ref{FigXrbImpurity}, respectively. With transfer reactions included,  urca cooling is now slightly reduced. This is due to the depletion by transfer reactions of $A=31$ and $A=98$ nuclei, preventing urca cooling via $^{31}$Na$\leftrightarrow ^{31}$Mg and $^{98}$Br$\leftrightarrow ^{98}$Kr at \density{8.7}{10} and \density{1.14}{11}, respectively. The $A=31$ nuclei are destroyed at \density{3.6}{10} via $^{31}$Mg($^{22}$O,$^{23}$O)$^{30}$Mg, while $A=98$ nuclei are destroyed at density \density{1.10}{11}, just before the urca cooling sets in, via $^{98}$Br($^{22}$O,$^{23}$O)$^{97}$Br. 

Another significant difference is the composition near the end of the calculation around \density{1.6}{12}. \citet{Lau2018} found that at that depth, the composition had consolidated into four dominant species (excluding free neutrons): $^{40}$Mg at neutron number $N=28$ (abundance $Y=\,$\EE{8.4}{-5}), $^{70}$Ca ($Y=\,$\EE{9.7}{-3}) at neutron number $N=50$, $^{116}$Se ($Y=\,$\EE{4.4}{-4}) at neutron number $N=82$, and $^{46}$Si ($Y=\,$\EE{4.4}{-4}) at neutron number $N=32$. With transfer reactions included, the composition is significantly more consolidated with only two dominant species left, $^{70}$Ca ($Y=\,$\EE{1.0}{-2}) and $^{46}$Si ($Y=\,$\EE{9.8}{-4}). $^{40}$Mg and $^{116}$Se are now only present in negligible amounts with $Y\approx\,$\EE{2}{-7}. The smaller variance in $Z$ results in a significantly reduced impurity parameter near the end of the calculation (Fig.~\ref{FigXrbImpurity}). 

The lack of $^{40}$Mg production is a direct result of the transfer reactions. Without transfer reactions, $^{40}$Mg is produced when $^{20}$O (created via electron captures from the initial $^{20}$Ne abundance in the rp-process ashes) reaches its electron capture threshold at around \density{1.33}{11}. A double electron capture then produces $^{20}$C. $^{40}$Mg is then produced by two production paths: directly via $^{20}$C$+^{20}$C$\rightarrow ^{40}$Mg pycnonuclear fusion, or via $^{20}$O$+^{20}$C$\rightarrow ^{40}$Si(2EC)$^{40}$Mg. However, when transfer reactions are included, $^{20}$O is already destroyed at \density{7.4}{10}, prior to formation of $^{40}$Mg. At that depth, a number of heavy nuclei, chiefly $^{105}$Sr, $^{69}$Fe, $^{92}$Se, $^{103}$Kr, $^{97}$Kr, and $^{99}$Kr transfer neutrons to $^{20}$O producing $^{21}$O. The neutron transfer reaction $^{21}$O($^{21}$O,$^{20}$O)$^{22}$O then converts two $^{21}$O into one $^{20}$O and one $^{22}$O. The net result is the conversion of $^{20}$O into $^{22}$O. 

\begin{figure}
\epsscale{1.15}
\plotone{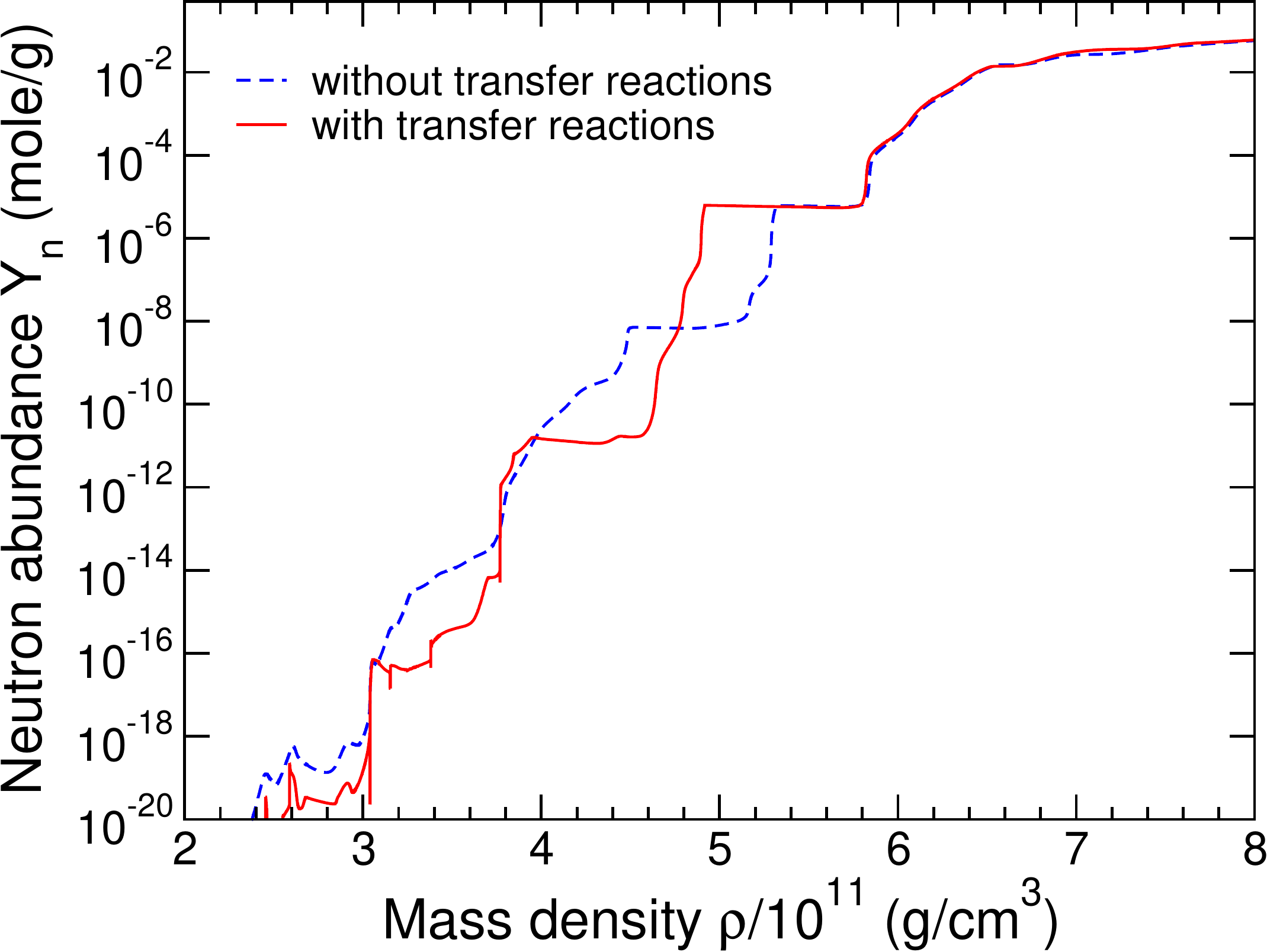}
\caption{\label{FigXRBYn} Free neutron abundance around neutron drip as a function of mass density for rp-process ashes with neutron transfer reactions (red, solid) and without (blue, dashed).}
\end{figure}

The reduced production of $^{116}$Se has a different explanation. The final amount of $^{116}$Se is determined by the competition of electron capture and neutron capture at the $^{108}$Se branchpoint reached at around \density{3.2}{11}. The reduced production of $^{116}$Se when transfer reactions are included is a result of the reduction of the free neutron abundance due to competition of neutron transfer with ($\gamma$,n) reactions. Fig.~\ref{FigXRBYn} shows the neutron abundance as a function of density with and without transfer reactions. While the onset of a significant free neutron abundance proceeds in a similar way, there are differences in neutron abundance in certain density ranges, which can be as large as two orders of magnitude. Coincidentally, at the depth of the critical branching at $^{108}$Se, neutron transfer reactions have reduced the free neutron abundance by more than an order of magnitude. As a result, the neutron capture branch leading to $^{116}$Se is drastically reduced. 

\section{Conclusions}
\begin{figure}
\epsscale{1.15}
\plotone{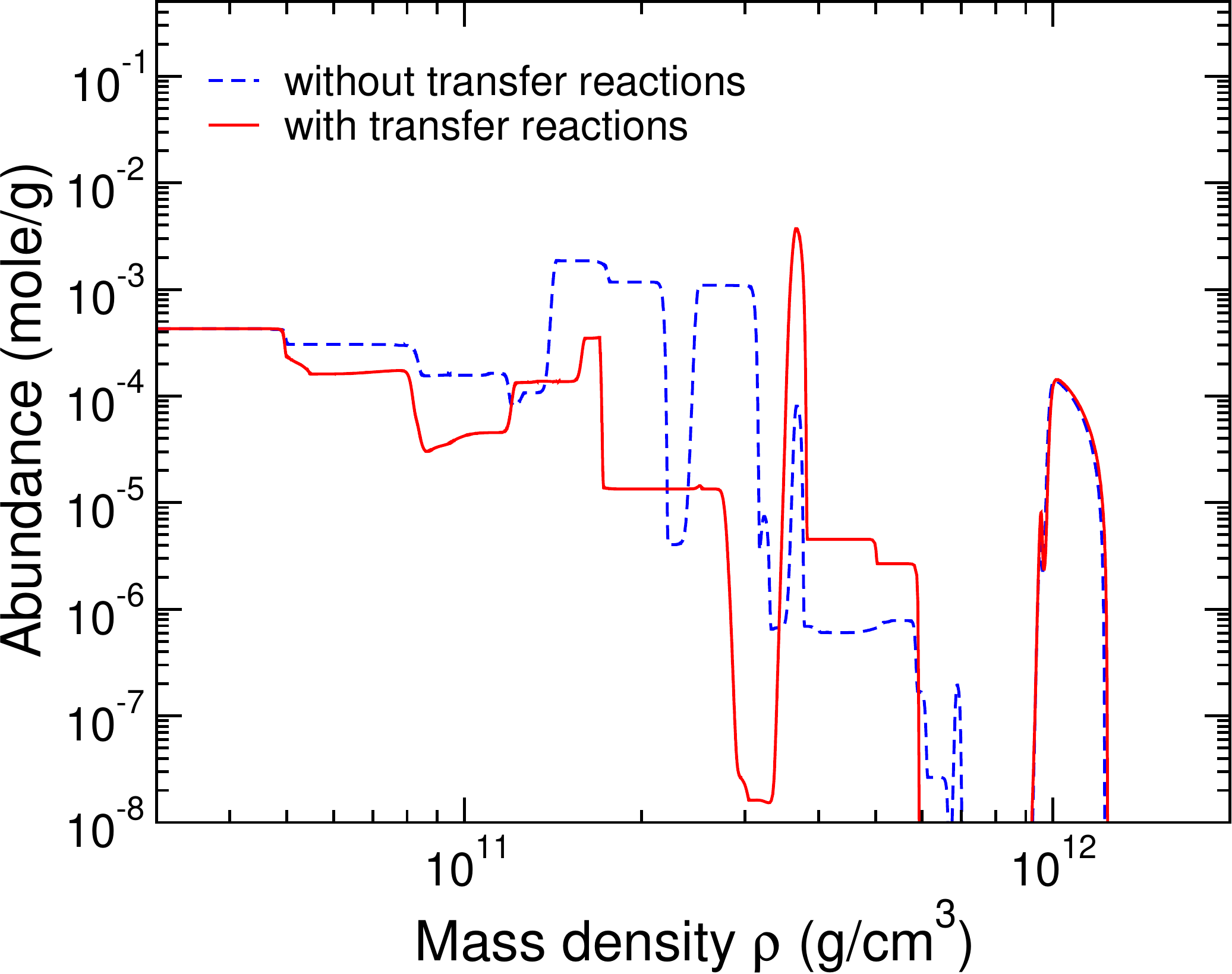}
\caption{\label{FigSBOddA} Summed abundances of all odd-$A$ nuclei for superburst ashes as a function of mass density with neutron transfer reactions (red, solid) and without (blue, dashed).}
\end{figure}

\begin{figure}
\epsscale{1.15}
\plotone{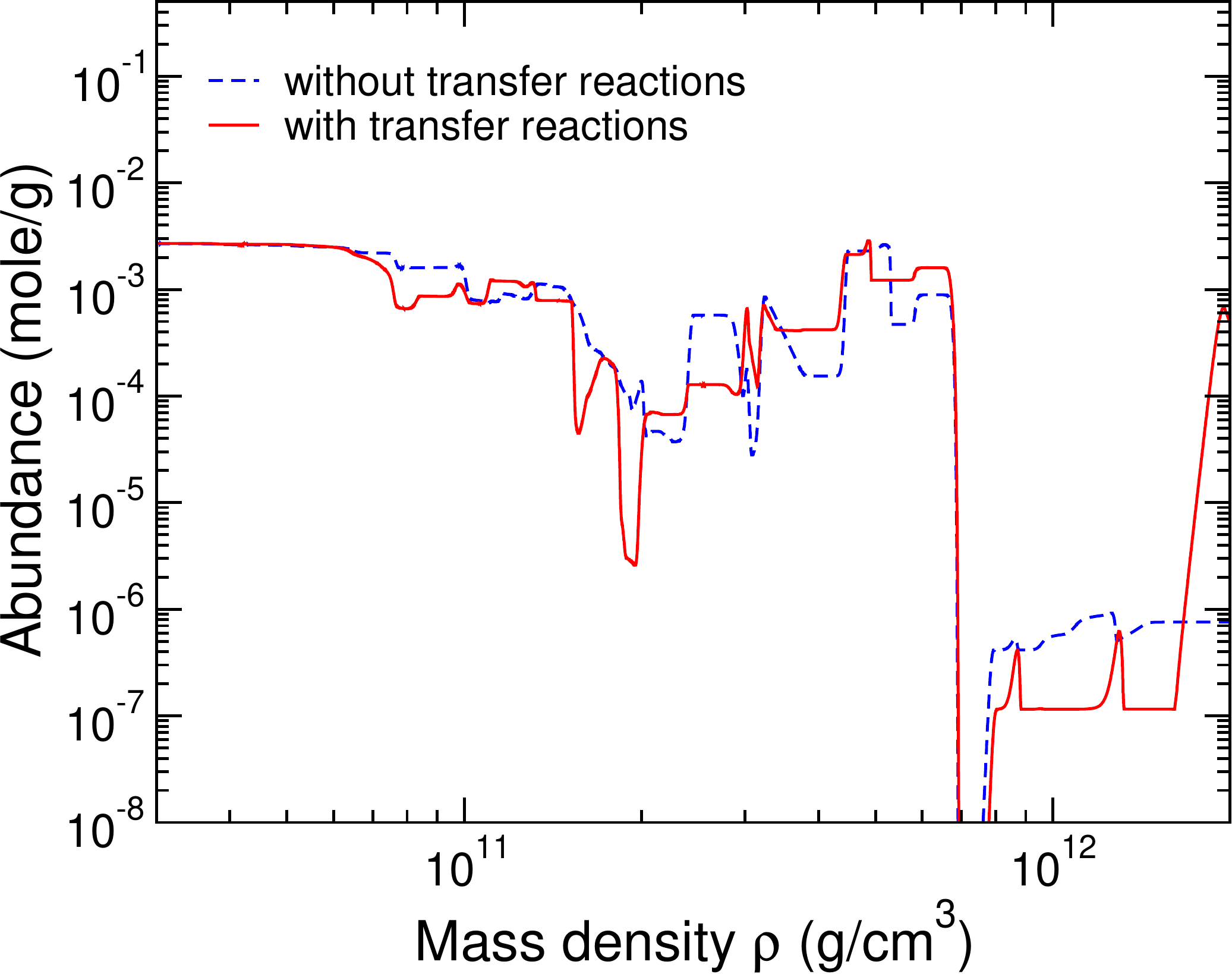}
\caption{\label{FigXRBOddA} Summed abundances of all odd-$A$ nuclei for  rp-process ashes as a function of mass density with neutron transfer reactions (red, solid) and without (blue, dashed).}
\end{figure}

We implemented neutron transfer reactions according to \citet{Chugunov2019} into the otherwise identical neutron star crust reaction network discussed in \citet{Lau2018}. We find that a large number of these reactions occur, and that they carry significant reaction flows. While there are significant differences, overall the resulting energy generation, cooling, free neutron abundances, and crust impurity are qualitatively similar to the results without neutron transfer reactions, and major conclusions do not change. In particular, the suggested early destruction of odd-$A$ nuclei and the resulting drastic suppression of crust urca cooling \citep{Chugunov2019} (as crust urca cooling occurs predominantly in odd-$A$ mass chains due to the single-step character of the electron capture sequence) is not as dramatic as expected. There is a complex interplay of reactions destroying odd-$A$ nuclei, and reactions producing them. While we do observe early destruction of some potentially urca-cooling odd-$A$ chains, for example $A=31$ in the case of rp-process ashes, confirming the prediction by \citet{Chugunov2019}, there are also cases where transfer reactions lead to an increased odd-$A$ abundance. In fact, for the case of superburst ashes, we find an increase in urca cooling due to a local increase in $A=29$ nuclei. The impact on odd-$A$ nuclei can also be seen in Fig.~\ref{FigSBOddA} and Fig.~\ref{FigXRBOddA}, which show the summed abundance of all odd-$A$ nuclei as a function of density. Indeed, transfer reactions tend to reduce the abundance of odd-$A$ nuclei, however, this is not the case everywhere, and there are also regions where they enhance odd-$A$ abundances. As nuclear urca processes in the crust occur in thin shells at specific densities, urca cooling can be either enhanced, or reduced by neutron transfer reactions.

Figs.~\ref{FigSBOddA} and \ref{FigXRBOddA} also show that even without neutron transfer reactions, similar trends occur in the odd-$A$ abundance with increasing density. At low densities, the odd-$A$ abundance is constant as $A$-preserving electron capture reactions and $\beta$-decays are the only reactions. At around \density{4.7}{10} the odd-$A$ (and accordingly the even-$A$) abundance begins to change, with or without neutron transfer reactions. In the model without neutron transfer reactions, ($\gamma$,n) reactions release free neutrons, which are then recaptured by other nuclei. Both, neutron release followed by capture, and neutron transfer, move neutrons from weakly neutron-bound nuclei to abundant, more neutron-bound nuclei, and both tend to reduce odd-$A$ nuclei as the double-step electron captures along even $A$ favor the more neutron-bound even-even nuclei. Both reaction types need to wait until such weakly neutron-bound nuclei are formed by electron capture, and therefore start at around the same depth. This explains the overall similarity between the calculations with and without neutron transfer reactions, as in the absence of neutron transfer reactions, neutron release and capture fulfill a similar role. 

However, the redistribution of neutrons by neutron transfer reactions is not the same as by neutron release and capture, and there are important differences in the results. Therefore, we confirm the conclusion from \citet{Chugunov2019} that neutron transfer reactions need to be taken into account in crust models for accurate predictions of urca cooling and composition. One key difference between neutron transfer and neutron emission with capture is that the former favors transfer to low $Z$ nuclei (which will tend to be low $A$ as well). This is due to the pair-correlation function (Eqn~\ref{eqn:pair}), as it matters how close the nuclei are in distance. Proximity is irrelevant for the neutron capture process as free neutrons can travel through the crystal lattice of the crust. As a consequence, neutron captures will predominantly occur on abundant high $A$ nuclei with large neutron capture cross sections, and all nuclear species present compete on equal footing for the available neutrons. 

Two examples where this plays out are the much reduced $A=57$ and $A=61$ abundances for superburst ashes, which in the absence of neutron transfer would be formed by neutron capture on the heaviest even-$A$ nuclei present. Another example is the conversion of $^{20}$O into $^{22}$O in the case of rp-process ashes by a two step transfer - first a heavier nucleus transfers a neutron to $^{20}$O forming $^{21}$O. Then the subsequent neutron transfer between two $^{21}$O nuclei produces $^{22}$O. Were the neutron in the heavy nucleus simply released, we would expect it to be captured by one of the many heavy nuclei present.  Therefore without neutron transfers, $^{20}$O survives and is at higher densities transformed to $^{40}$Mg. 

Another important impact of neutron transfer reactions we find is a reduction in free neutron abundance. As neutron transfers compete with ($\gamma$,n) neutron release for removing neutrons from weakly neutron bound nuclei, the inclusion of neutron transfers reduces the number of ($\gamma$,n) reactions. We find that this has a significant effect on the crust composition for both  superburst and rp-process ashes. This is due to shifts in (n,$\gamma$)$-$($\gamma$,n) equilibrium cycles, and again due to the fact that neutron transfers tend to distribute more neutrons to lighter nuclei. The lower free neutron abundance at the depth of the critical $^{108}$Se branchpoint prevents formation of significant amounts of $N=82$ $^{116}$Se in the model with rp-process ashes.  

Arguably, one of the most important changes due to neutron transfer reactions relates to the surprising finding of increased crust impurity at relatively high densities around \density{1.6}{12} in \citet{Lau2018} for rp-process ashes. This is a result of nuclear structure effects that lead to the lock-in of abundance at the classical $N=28, 50, 82$ shell closures. With the suppression of $^{40}$Mg and $^{116}$Se production due to neutron transfer reactions discussed above, the built up of abundance at $N=28$ and $N=82$ is very small and the peak in impurity is much reduced. However, there are many other nuclear uncertainties. For example the free neutron abundance at the $^{108}$Se branchpoint will depend sensitively on the initial abundance distribution that directly effects the release of free neutrons. There are many nuclear uncertainties in the rp-process \citep{Parikh2008,Cyburt2016}, in particular in the production of the odd-$A$ nuclei \citep{Merz2021} that largely determines neutron emission. In addition, the nuclear structure effects that lead to a concentration and a ``lock-in" of abundances in a small number of nuclei near the neutron drip line due to the impact of nuclear structure on nuclear masses are not well understood and experimental data are lacking. It is therefore too early to draw any final conclusions on the impurity around neutron drip. 

The neutron transfer reaction rates employed thus far are rough approximations. Improvements would include using theoretical estimates for nuclear level densities, more exact nuclear wave functions and mean-force potentials, and employing numerical integration rather than more convenient approximations.

\acknowledgments
We acknowledge stimulating discussions with L. Bildsten, D. Yakovlev, A. Cumming, and within the JINA-CEE crust working group. This work was supported in part by the US National Science Foundation under grant PHY-1430152 (JINA Center for the Evolution of the Elements).
Support by the US National Science Foundation is acknowledged by E.F.B. under grant 80NSSC20K0503 from NASA, and by H.S. under grant PHY-1102511 and PHY-1913554. Support by the US Department of Energy, Office of Science is acknowledged by Z.M. under Award No. DE-FG02-88ER40387 and DE-SC0019042.
\bibliography{hsref_v6}

\begin{thebibliography}{}
\expandafter\ifx\csname natexlab\endcsname\relax\def\natexlab#1{#1}\fi

\bibitem[{{Arnett}(1996)}]{Arnett1996}
{Arnett}, D. 1996, {Supernovae and Nucleosynthesis} (Princeton University
  Press)

\bibitem[{{Brown} \& {Cumming}(2009)}]{Brown2009}
{Brown}, E.~F., \& {Cumming}, A. 2009, \apj, 698, 1020

\bibitem[{{Cackett} {et~al.}(2006){Cackett}, {Wijnands}, {Linares}, {Miller},
  {Homan}, \& {Lewin}}]{Cackett2006}
{Cackett}, E.~M., {Wijnands}, R., {Linares}, M., {et~al.} 2006, \mnras, 372,
  479

\bibitem[{{Chamel} {et~al.}(2020){Chamel}, {Fantina}, {Zdunik}, \&
  {Haensel}}]{Chamel2020}
{Chamel}, N., {Fantina}, A.~F., {Zdunik}, J.~L., \& {Haensel}, P. 2020, \prc,
  102, 015804

\bibitem[{{Chugunov}(2019)}]{Chugunov2019}
{Chugunov}, A.~I. 2019, \mnras, 483, L47

\bibitem[{{Cyburt} {et~al.}(2016){Cyburt}, {Amthor}, {Heger}, {Johnson},
  {Keek}, {Meisel}, {Schatz}, \& {Smith}}]{Cyburt2016}
{Cyburt}, R.~H., {Amthor}, A.~M., {Heger}, A., {et~al.} 2016, \apj, 830, 55

\bibitem[{{Degenaar} {et~al.}(2011){Degenaar}, {Brown}, \&
  {Wijnands}}]{Degenaar2011}
{Degenaar}, N., {Brown}, E.~F., \& {Wijnands}, R. 2011, \mnras, 418, L152

\bibitem[{{Degenaar} {et~al.}(2013){Degenaar}, {Wijnands}, {Brown},
  {Altamirano}, {Cackett}, {Fridriksson}, {Homan}, {Heinke}, {Miller},
  {Pooley}, \& {Sivakoff}}]{Degenaar2013}
{Degenaar}, N., {Wijnands}, R., {Brown}, E.~F., {et~al.} 2013, \apj, 775, 48

\bibitem[{{Degenaar} {et~al.}(2015){Degenaar}, {Wijnands}, {Bahramian},
  {Sivakoff}, {Heinke}, {Brown}, {Fridriksson}, {Homan}, {Cackett}, {Cumming},
  {Miller}, {Altamirano}, \& {Pooley}}]{Degenaar2015}
{Degenaar}, N., {Wijnands}, R., {Bahramian}, A., {et~al.} 2015, \mnras, 451,
  2071

\bibitem[{{Deibel} {et~al.}(2015){Deibel}, {Cumming}, {Brown}, \&
  {Page}}]{Deibel2015}
{Deibel}, A., {Cumming}, A., {Brown}, E.~F., \& {Page}, D. 2015, \apjl, 809,
  L31

\bibitem[{{Deibel} {et~al.}(2017){Deibel}, {Cumming}, {Brown}, \&
  {Reddy}}]{Deibel2017}
{Deibel}, A., {Cumming}, A., {Brown}, E.~F., \& {Reddy}, S. 2017, \apj, 839, 95

\bibitem[{{Deibel} {et~al.}(2016){Deibel}, {Meisel}, {Schatz}, {Brown}, \&
  {Cumming}}]{Deibel2016}
{Deibel}, A., {Meisel}, Z., {Schatz}, H., {Brown}, E.~F., \& {Cumming}, A.
  2016, \apj, 831, 13

\bibitem[{{Gupta} {et~al.}(2007){Gupta}, {Brown}, {Schatz}, {M{\"o}ller}, \&
  {Kratz}}]{Gupta2007}
{Gupta}, S., {Brown}, E.~F., {Schatz}, H., {M{\"o}ller}, P., \& {Kratz}, K.-L.
  2007, \apj, 662, 1188

\bibitem[{{Gupta} {et~al.}(2008){Gupta}, {Kawano}, \& {M{\"o}ller}}]{Gupta2008}
{Gupta}, S.~S., {Kawano}, T., \& {M{\"o}ller}, P. 2008, Phys. Rev. Lett., 101,
  231101

\bibitem[{{Haensel} \& {Zdunik}(1990)}]{Haensel1990}
{Haensel}, P., \& {Zdunik}, J.~L. 1990, \AsAs, 227, 431

\bibitem[{Haensel \& Zdunik(2008)}]{Haensel2008}
Haensel, P., \& Zdunik, J.~L. 2008, \AsAs, 480, 459

\bibitem[{{Horowitz} {et~al.}(2015){Horowitz}, {Berry}, {Briggs}, {Caplan},
  {Cumming}, \& {Schneider}}]{Horowitz2015}
{Horowitz}, C.~J., {Berry}, D.~K., {Briggs}, C.~M., {et~al.} 2015, Phys. Rev.
  Lett., 114, 031102

\bibitem[{Keek {et~al.}(2012)Keek, Heger, \& in't Zand}]{Keek2012}
Keek, L., Heger, A., \& in't Zand, J. J.~M. 2012, \apj, 752, 150

\bibitem[{Lau {et~al.}(2018)Lau, Beard, Gupta, Schatz, Afanasjev, Brown,
  Deibel, Gasques, Hitt, Hix, Keek, M\"{o}ller, Shternin, Steiner, Wiescher, \&
  Xu}]{Lau2018}
Lau, R., Beard, M., Gupta, S.~S., {et~al.} 2018, \apj, 859, 62

\bibitem[{{Meisel} {et~al.}(2018){Meisel}, {Deibel}, {Keek}, {Shternin}, \&
  {Elfritz}}]{Meisel2018}
{Meisel}, Z., {Deibel}, A., {Keek}, L., {Shternin}, P., \& {Elfritz}, J. 2018,
  Journal of Physics G Nuclear Physics, 45, 093001

\bibitem[{{Merritt} {et~al.}(2016){Merritt}, {Cackett}, {Brown}, {Page},
  {Cumming}, {Degenaar}, {Deibel}, {Homan}, {Miller}, \&
  {Wijnands}}]{Merritt2016}
{Merritt}, R.~L., {Cackett}, E.~M., {Brown}, E.~F., {et~al.} 2016, \apj, 833,
  186

\bibitem[{{Merz} \& {Meisel}(2021)}]{Merz2021}
{Merz}, G., \& {Meisel}, Z. 2021, \mnras, 500, 2958

\bibitem[{{Ootes} {et~al.}(2016){Ootes}, {Page}, {Wijnands}, \&
  {Degenaar}}]{Ootes2016}
{Ootes}, L.~S., {Page}, D., {Wijnands}, R., \& {Degenaar}, N. 2016, \mnras,
  461, 4400

\bibitem[{{Page} \& {Reddy}(2013)}]{Page2013}
{Page}, D., \& {Reddy}, S. 2013, Phys. Rev. Lett., 111, 241102

\bibitem[{{Parikh} {et~al.}(2008){Parikh}, {Jos{\'e}}, {Moreno}, \&
  {Iliadis}}]{Parikh2008}
{Parikh}, A., {Jos{\'e}}, J., {Moreno}, F., \& {Iliadis}, C. 2008, \apjs, 178,
  110

\bibitem[{{Parikh} {et~al.}(2019){Parikh}, {Wijnands}, {Ootes}, {Page},
  {Degenaar}, {Bahramian}, {Brown}, {Cackett}, {Cumming}, {Heinke}, {Homan},
  {Rouco Escorial}, \& {Wijngaarden}}]{Parikh2019}
{Parikh}, A.~S., {Wijnands}, R., {Ootes}, L.~S., {et~al.} 2019, \aap, 624, A84

\bibitem[{{Potekhin} \& {Chabrier}(2021)}]{Potekhin2021}
{Potekhin}, A.~Y., \& {Chabrier}, G. 2021, \aap, 645, A102

\bibitem[{{Rutledge} {et~al.}(2002){Rutledge}, {Bildsten}, {Brown}, {Pavlov},
  {Zavlin}, \& {Ushomirsky}}]{Rutledge2002}
{Rutledge}, R.~E., {Bildsten}, L., {Brown}, E.~F., {et~al.} 2002, \apj, 580,
  413

\bibitem[{{Sato}(1979)}]{Sato1979}
{Sato}, K. 1979, Prog. Theor. Phys., 62, 957

\bibitem[{Schatz {et~al.}(2001)Schatz, Aprahamian, Barnard, Bildsten, Cumming,
  Ouellette, Rauscher, Thielemann, \& Wiescher}]{Schatz2001}
Schatz, H., Aprahamian, A., Barnard, V., {et~al.} 2001, Phys. Rev. Lett., 86,
  3471

\bibitem[{{Schatz} {et~al.}(2014){Schatz}, {Gupta}, {M{\"o}ller}, {Beard},
  {Brown}, {Deibel}, {Gasques}, {Hix}, {Keek}, {Lau}, {Steiner}, \&
  {Wiescher}}]{Schatz2014}
{Schatz}, H., {Gupta}, S., {M{\"o}ller}, P., {et~al.} 2014, \nat, 505, 62

\bibitem[{{Shchechilin} \& {Chugunov}(2019)}]{Shchechilin2019}
{Shchechilin}, N.~N., \& {Chugunov}, A.~I. 2019, \mnras, 490, 3454

\bibitem[{{Shchechilin} {et~al.}(2021){Shchechilin}, {Gusakov}, \&
  {Chugunov}}]{Shchechilin2021}
{Shchechilin}, N.~N., {Gusakov}, M.~E., \& {Chugunov}, A.~I. 2021, arXiv
  e-prints, arXiv:2105.01991

\bibitem[{{Shternin} {et~al.}(2007){Shternin}, {Yakovlev}, {Haensel}, \&
  {Potekhin}}]{Shternin2007}
{Shternin}, P.~S., {Yakovlev}, D.~G., {Haensel}, P., \& {Potekhin}, A.~Y. 2007,
  \mnras, 382, L43

\bibitem[{{Turlione} {et~al.}(2015){Turlione}, {Aguilera}, \&
  {Pons}}]{Turlione2015}
{Turlione}, A., {Aguilera}, D.~N., \& {Pons}, J.~A. 2015, \aap, 577, A5

\bibitem[{{Waterhouse} {et~al.}(2016){Waterhouse}, {Degenaar}, {Wijnands},
  {Brown}, {Miller}, {Altamirano}, \& {Linares}}]{Waterhouse2016}
{Waterhouse}, A.~C., {Degenaar}, N., {Wijnands}, R., {et~al.} 2016, \mnras,
  456, 4001

\end{thebibliography}

\end{document}